\documentclass
[groupedaddress,doublelines,10pt,twocolumn,final,prl,showpacs]{revtex4}%
\usepackage{amssymb}
\usepackage{amsmath}
\usepackage{amsfonts}
\usepackage{graphicx}%
\providecommand{\U}[1]{\protect\rule{.1in}{.1in}}

\begin{document}
\title{Quantum Neural Network and Soft Quantum Computing}
\author{Zeng-Bing Chen}
\affiliation{National Laboratory of Solid State Microstructures, School of Physics, Nanjing
University, Nanjing 210093, China}
\date{\today}

\begin{abstract}
A new paradigm of quantum computing, namely, soft quantum computing, is
proposed for nonclassical computation using real world quantum systems with
naturally occurring environment-induced decoherence and dissipation. As a
specific example of soft quantum computing, we suggest a quantum neural
network, where the neurons connect pairwise via the \textquotedblleft controlled Kraus
operations\textquotedblright, hoping to pave an easier and more realistic way to quantum artificial
intelligence and even to better understanding certain functioning of the
human brain. Our quantum neuron model mimics as much as possible the realistic neurons and meanwhile, uses quantum laws for processing information. The quantum features of the noisy neural network
are uncovered by the presence of quantum discord and by non-commutability of
quantum operations. We believe that our model puts quantum computing into a
wider context and inspires the hope to build a soft quantum computer much
earlier than the standard one.

\end{abstract}

\maketitle

Conventional (hard) computing is characterized by precision, certainty, and
rigor. By contrast, \textquotedblleft soft computing\textquotedblright%
\ \cite{Zadeh,Konar}\ is an approach to computing, which mimics the human
thinking to learn and reason in an environment of imprecision, uncertainty,
and partial truth, yet to achieve tractability, robustness, and low solution
cost. The applications of soft computing cover various areas of fuzzy logic,
neural networks, evolutionary computing, rough sets, and other similar
techniques to address real world complexities. Another paradigm shift in
computing is the exploitation of quantum computing, first introduced by
Feynman in 1982. In its circuit-based model, quantum computing \cite{Preskill}
consists of initializing input qubits, manipulating those qubits with a fixed
sequence of quantum logic gates programmed by a specific quantum algorithm,
and finally measuring all the output qubits. While recent years have witnessed
remarkable progresses on physical implementations of quantum computing, there
still exist a number of significant challenges in building a large-scale
quantum computer. The main technical challenges include the high-precision
initialization and readout of qubits well isolated from environment,
high-precision quantum logic gates, and scalability of the physical qubits to
a large-scale quantum computing device. Quantum error correction can correct
errors occurred during computing, thereby allowing to overcome the decoherence
effects caused by the noisy environment or faulty quantum logic gates.
However, the required error rate for each gate is extremely low, typically
$10^{-4}$, which is very hard to fulfill for current quantum computing
systems; for a topological approach to quantum computing with much better
fault tolerance, see Refs.~\cite{topo,topo-exp}.

A realistic quantum system is always characterized by non-unitary, faulty
evolutions and coupled with the noisy and dissipative environment. The real
system complexities in quantum domain call for a new paradigm of quantum
computing, namely, soft (or natural) quantum computing, aiming at nonclassical
computation using real world quantum systems with naturally occurring
environment-induced decoherence and dissipation. Thus, by its very definition,
soft quantum computing deals with nontrivial (i.e., classically intractable)
quantum computing under the conditions of noisy and faulty quantum evolutions
and measurements, while being tolerant to those effects that are detrimental
for current quantum computing paradigm. Thus, unlike conventional (hard)
quantum computing, soft quantum computing does not aim at universal
computation, but certain specific computational tasks in an open quantum system.

The most important question that soft quantum computing attempts to address is
whether or not this new paradigm shift in computing could help in a better
understanding of certain functioning of the human brain (\textquotedblleft
quantum artificial intelligence\textquotedblright). Thus, in this work we
propose a quantum neural network model as an illustration of soft quantum
computing. The belief behind the proposal stems from the recent exciting
discoveries on possible quantum mechanical effects in biological systems (for
a review, see Ref.~\cite{q-bio}). If photosynthesis \cite{photosyn} and avian
navigation \cite{avian} can make use of quantum effects in certain manner, why
the human brain---the most marvellous biological device---cannot utilize
quantum laws to enhance its computing and reasoning power? Along this line,
models of quantum cognition based on, e.g., neuronal microtubules
\cite{microtubule} and nuclear spins \cite{med-hyp,q-cog-spin} were
envisioned. Here, rather than focusing on physical implementations of quantum
functioning of the brain, we are interested in the mathematical model of
quantum neural network which works as a soft quantum computer. For a summary
of existing quantum neural network models, see Ref.~\cite{quantumNN-rev}.

Mathematically, soft quantum computing starts with $n$ two-level quantum
systems (qubits) coupled with their surrounding environment, and as such the
initial state of the $n$ qubits is a mixed state $\rho_{12...n}$ in the
computational basis $\left\vert 0\right\rangle $ and $\left\vert
1\right\rangle $. The evolution for such a soft quantum computer during
computing is described by a completely positive Kraus map (denoted by a
superoperator $\mathcal{O}$)
\begin{equation}
\mathcal{O}(\rho_{12...n})=\sum_{k}\hat{E}_{k}(t)\rho_{12...n}\hat{E}%
_{k}^{\dagger}(t). \label{kraus}%
\end{equation}
Here the quantum operations $\hat{E}_{k}$, which might be time-dependent,
satisfy $\sum_{k}\hat{E}_{k}(t)\hat{E}_{k}^{\dagger}(t)\leq I$, where the
equality holds for a trace-preserving map. Important examples of the
trace-preserving operations are projective measurements, unitary evolutions
and partial tracing. After the noisy evolution, quantum measurement in the
computational basis completes the soft quantum computing process. Below, we
model neurons as noisy qubits; the network of such noisy qubits under noisy
evolution and measurement is a particular model of soft quantum computing.

The human brain can be regarded as a neural network \cite{Konar,Kumar}
organized in a formidably sophisticated structure and has a huge number
($\sim10^{11}$) of neurons. A drastically simplified drawing of a neuron is
shown in Fig.~1a. A real neuron, as we now understand it, integrates hundreds
or thousands of impinging signals through its dendrites. These signals result
in a change of the internal cell potential in a complex pattern that depends
on the excitatory or inhibitory nature of the synapses where the signals
impinge on the cell body. The neuron outputs a signal through its axon to
another neuron for processing in the form of an action potential only when its
internal potential exceeds certain threshold. Note that a neuron has only a
single axon, but a set of dendrites forming a tree-like structure.

\begin{figure}[ptb]%
\centering
\includegraphics[
width=2.8236in
]%
{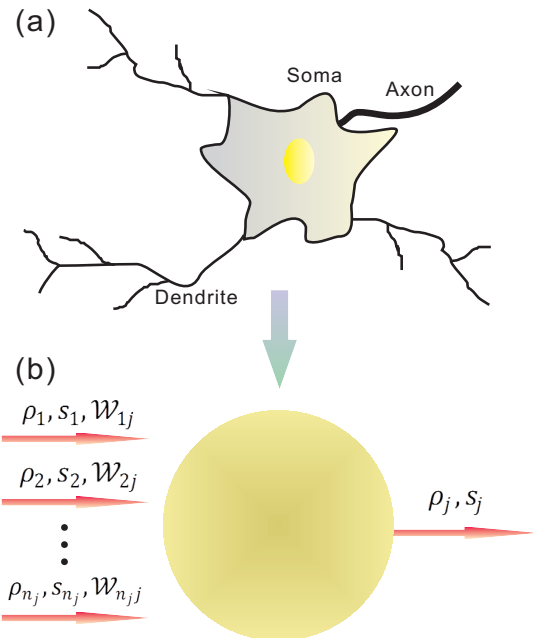}%
\caption{A drastically simplified drawing of a neuron (a) and the soft quantum
neuron model (b). In (a), the neuron integrates hundreds or thousands of
impinging signals through its dendrites. After processing by the cell body,
the neuron outputs a signal through its axon to another neuron for processing
in the form of an action potential when its internal potential exceeds certain
threshold.}%
\label{figqnn}%
\end{figure}

Instead of modeling the neurons as in the conventional neural network models,
or using the concepts borrowed from hard quantum computing
\cite{quantumNN-rev}, we model the $j$th neuron within the neural network as a
basic unit of a soft quantum computer; the network can have any network
architecture, which we do not specify below. As shown in Fig.~1b, in our
quantum neural model each neuron has $n_{j}$ inputs $s_{i}$ $(i=1,2,...,n_{j}%
$) from possibly $n_{j}$ other neurons, each of which \textquotedblleft
interacts\textquotedblright\ with the $j$th neuron. The interaction is
characterized by a connection superoperator $\mathcal{W}_{ij}(t)$ instead of a
connection matrix (or weighted pathways) $w_{ij}$ in the conventional neural
networks \cite{Konar,Kumar}. Here we have shown the time-dependence of
$\mathcal{W}_{ij}(t)$ explicitly. Then we assume that the $j$th neuron is a
logic qubit, consisting of $n_{j}$ physical qubits. The candidate of these
physical qubits could be neuron's ion channels whose two conducting states
(open and closed states) qualify them as the two-state systems. The detailed
microscopic justification of physical qubits, which will be considered in
future, is of course meaningful, but irrelevant here for our mathematical
formulation of the quantum neuron model. These $n_{j}$ physical qubits are
operated by $n_{j}$ inputs via the connection superoperator $\mathcal{W}%
_{ij}(t)$. As in a real neuron integrating impinging signals and outputing a
signal when the internal potential of the neuron exceeds certain threshold,
here the resulting collective state of the $n_{j}$ physical qubits represents
the integration of impinging signals and determines the output state of the
logic qubit. Thus, \textit{we model the quantum neuron at each network node
simultaneously as }$n_{j}$\textit{ physical qubits and as a logic qubit} such
that it can interact with $n_{j}$ impinging signals and output a
\textit{single} qubit state. In this way, the neuron model proposed above
mimics as much as possible the realistic neurons and meanwhile, uses quantum
laws for processing information.

The logic qubit is initially described by a density matrix $\rho_{j}^{in}$. By
modeling each neuron as a logic qubit, we only consider binary signals,
namely, $s_{i}=0$ or $1$. Whenever $s_{i}=1$, corresponding to the case of a
signal input, a quantum operation (here, a superoperator) $\mathcal{W}_{ij}$
will be acted upon $\rho_{j}^{in}$, while for $s_{i}=0$ (zero input signal),
nothing will happen to $\rho_{j}^{in}$. Obviously, such a conditional
operation, called the \textquotedblleft controlled Kraus
operation\textquotedblright\ hereafter, is the generalization of the two-qubit
controlled unitary operations in hard quantum computing. Meanwhile, each input
neuron itself with input $s_{i}$ is also a logic qubit in a mixed state
$\rho_{i}^{in}$. As a result, the evolution of the whole system from the
initial state $\bigotimes\nolimits_{i=1}^{n_{j}}\rho_{i}^{in}\otimes\rho
_{j}^{in}$ reads
\begin{align}
\rho_{\left\{  i\right\}  j}^{out}  &  =\mathcal{T}\bigotimes\nolimits_{i=1}%
^{n_{j}}\mathcal{O}_{ij}(\rho_{i}^{in}\otimes\rho_{j}^{in})\nonumber\\
&  \equiv\mathcal{T}\bigotimes\nolimits_{i=1}^{n_{j}}[(\mathcal{P}_{\left\vert
0\right\rangle _{i}}\otimes\hat{I}_{j}+\mathcal{P}_{\left\vert 1\right\rangle
_{i}}\otimes\mathcal{W}_{ij}(t)](\rho_{i}^{in}\otimes\rho_{j}^{in}).
\label{evo}%
\end{align}
Here the \textquotedblleft superprojectors\textquotedblright\ $\mathcal{P}%
_{\left\vert s\right\rangle }$ are defined by $\mathcal{P}_{\left\vert
s\right\rangle }\rho=\left\vert s\right\rangle \left\langle s\right\vert
\rho\left\vert s\right\rangle \left\langle s\right\vert $, $\hat{I}$ is the
identity operator, and $\mathcal{T}$ represents the time-ordering. Generally
speaking, all $\mathcal{W}_{ij}(t)$ are acted upon the target neuron with
specific temporal patterns and as such, the time-ordering of these actions is
important as different $\mathcal{W}_{ij}(t)$ might be noncommutative. Note
that, in addition to the conditional dynamical evolution in Eq.~(\ref{evo}),
each quantum neuron can also be subject to local noisy operations.

Here some remarks are in order. A neuron of the human brain normally has about
$10^{4}$ synapses on average. It is this large number of input neurons that
activate the target neuron. According to modern neuroscience, the connection
matrix $w_{ij}$ is the mathematical abstraction of the synaptic efficacies of
the inter-neuron synapses, while impinging signals mathematically represent
action potentials. In classical neuron model, the activation of a real neuron
is then determined by the internal integrated cell potential, or
mathematically by the signal function and the internal firing threshold. For
real neurons, as action potentials and cell potentials change much faster than
changes in synaptic efficacies, neuron activations are a fast process, while
the changes of synaptic efficacies are a slow one. In classical neural
networks, the change of synaptic efficacies is implemented by a
\textquotedblleft learning algorithm\textquotedblright\ \cite{Konar,Kumar}%
\ during a learning process in response to stable patterns of activity.

However, in the context of our current quantum neuron model, the above remarks
should be looked from a different angle. Here a learning process is realized
by a \textquotedblleft quantum learning algorithm\textquotedblright\ (see,
e.g., Ref.~\cite{QLA-rev} for a review), i.e., a specific temporal pattern of
quantum operations $\mathcal{W}_{ij}(t)$ depending on the states of input
neurons. By using various quantum learning algorithms, the quantum neuron
model opens a door to quantum manipulations of real neurons, artificial
intelligence, and ultimately, the human brain. By sharp contrast, the existing
quantum neural networks \cite{quantumNN-rev} model neurons as qubits in pure
states and connection weights as usual quantum logic gates. However, the human
brain is a highly open and decohering system. It is very unlikely to model the
human brain as a well isolated quantum system. In our model, \textit{the human
brain is neither a classical computer (soft or hard) nor a standard quantum
computer, but something in between, namely, a soft quantum computer}. By
modeling neuron nodes as qubits in mixed states and their connections as
connection superoperators, we are left to see to what extend our model could
simulate and understand the real functioning of the human brain.

We used the superprojectors\ $\mathcal{P}_{\left\vert s\right\rangle }$ in
Eq.~(\ref{evo}) to select the computational basis as the input to the $j$th
neuron. In doing so, we have actually implicitly assumed certain
environment-induced decoherence \cite{Zurek} naturally occurring in the cell
body to choose the preferred basis, namely, the computational basis. To be
consistent for our quantum neuron model, we have to determine the output
$s_{j}$ of the $j$th neuron in the same way. The final state of the $j$th
neuron can be obtained by tracing out all the input states, namely, $\rho
_{j}^{out}=\mathrm{tr}_{\left\{  i\right\}  }\rho_{\left\{  i\right\}
j}^{out}=\mathcal{T}\prod_{i=1}^{n_{j}}[p_{i}\hat{I}_{j}+(1-p_{i}%
)\mathcal{W}_{ij}(t)]\rho_{j}^{in}$, where $p_{i}\equiv p_{i}(0)=\mathrm{tr}%
(\left\vert 0\right\rangle _{i}\left\langle 0\right\vert \rho_{i}^{in})$ is
the probability of $\left\vert 0\right\rangle _{i}$ within $\rho_{i}^{in}$.
The output state is then
\begin{equation}
\rho_{j}=(\mathcal{P}_{\left\vert 0\right\rangle _{j}}+\mathcal{P}_{\left\vert
1\right\rangle _{j}})\rho_{j}^{out}. \label{roj}%
\end{equation}
In other words, the output signal $s_{j}$ of the $j$th neuron is
\begin{equation}
s_{j}=\left\{
\begin{array}
[c]{ll}%
0\text{ \ \ } & \text{with probability }p_{j}(0)\\
1\text{ \ \ } & \text{with probability }1-p_{j}(0)
\end{array}
\right.  \label{sj}%
\end{equation}
This completes the specification of our quantum neural network model as a soft
quantum computer. Interestingly, the above prescriptions on the input and
output states in the preferred computational basis tacitly assumed each neuron
in our model as a quantum self-measuring meter. This eliminates the ambiguity
of designating signal functions of various forms in classical neural networks
\cite{Kumar}.

Physically, one could regard each input dendrite as a quantum channel
\cite{Preskill}, which carries out a quantum manipulation $\mathcal{W}_{ij}$
if and only if $s_{i}=1$. The depolarizing channel can be represented by
$\mathcal{W}_{dp}(\rho)=\sum_{\mu}M_{\mu}\rho M_{\mu}^{\dag}$ with
$M_{0}=\sqrt{1-p}\hat{I}\ $and $M_{k}=\sqrt{\frac{p}{3}}\sigma_{k}$
($k=1,2,3$; $0\leq p\leq1$), where $\sigma_{k}$ are the Pauli operators. For a
phase-damping channel $\mathcal{W}_{pd}(\rho)=\sum_{\mu}M_{\mu}\rho M_{\mu
}^{\dag}$, where $M_{0}=\sqrt{1-p}\hat{I}$, $M_{1}=\frac{\sqrt{p}}{2}%
(1+\sigma_{3})$,\ and $M_{2}=\frac{\sqrt{p}}{2}(1-\sigma_{3})$, while for an
amplitude-damping channel $\mathcal{W}_{ad}(\rho)=\sum_{\mu}M_{\mu}\rho
M_{\mu}^{\dag}$ one has two Kraus operators $M_{0}=\left\vert 0\right\rangle
\left\langle 0\right\vert +\sqrt{1-p}\left\vert 1\right\rangle \left\langle
1\right\vert \ $and$\ M_{1}=\sqrt{p}\left\vert 0\right\rangle \left\langle
1\right\vert $.

Now it is ready to discuss some key features of the current quantum neural
network or soft quantum computing. First we consider the simplest two-neuron
case. For the two neurons in the initial states\ $\rho_{1}^{in}=p_{1}%
\left\vert 0\right\rangle _{1}\left\langle 0\right\vert +(1-p_{1})\left\vert
1\right\rangle _{1}\left\langle 1\right\vert $\ ($p_{1}\neq0,1$) and $\rho
_{2}^{in}$, the action of a controlled Kraus operation $\mathcal{O}_{12}$
results in the final state
\begin{align}
\rho_{12}^{out}  &  =(\mathcal{P}_{\left\vert 0\right\rangle _{1}}\otimes
\hat{I}_{2}+\mathcal{P}_{\left\vert 1\right\rangle _{1}}\otimes\mathcal{W}%
_{12})(\rho_{1}^{in}\otimes\rho_{2}^{in})\nonumber\\
&  =p_{1}\left\vert 0\right\rangle _{1}\left\langle 0\right\vert \otimes
\rho_{2}^{in}+(1-p_{1})\left\vert 1\right\rangle _{1}\left\langle 1\right\vert
\otimes\mathcal{W}_{12}(\rho_{2}^{in}), \label{tn}%
\end{align}
where $\mathcal{W}_{12}$ represents a specific quantum channel. Is $\rho
_{12}^{out}$ nonclassically (or, quantum) correlated? Quantum correlations, if
any, of $\rho_{12}^{out}$ can be quantified by the quantum discord
\cite{discord}. Any bipartite state is called fully classically correlated if
it is of the form \cite{disc-adPRL}
\begin{equation}
\rho_{12}=\sum_{i,j}p_{ij}\left\vert i\right\rangle _{1}\left\langle
i\right\vert \otimes\left\vert j\right\rangle _{2}\left\langle j\right\vert ;
\label{clc}%
\end{equation}
otherwise, it is quantum correlated. Here $\left\vert i\right\rangle _{1}$ and
$\left\vert j\right\rangle _{2}$ are the orthonormal bases of the two parties,
with nonnegative probabilities $p_{ij}$.

Obviously, for the two-neuron state $\rho_{12}^{out}$ in Eq.~(\ref{tn}) the
first neuron becomes correlated with nonorthogonal states of the second neuron
as far as $\mathcal{W}_{12}(\rho_{2}^{in})$ and $\rho_{2}^{in}$ are
nonorthogonal \cite{Brukner,disc-adPRA,disc-adPRLe}, namely, $\rho_{12}^{out}$
has quantum correlations. In particular, Refs.~\cite{disc-adPRA,disc-adPRLe}
show the creation of discord, from classically correlated two-qubit states, by
applying an amplitude-damping process only on one of the qubits; for the
phase-damping process, see Ref.~\cite{disc-adPRL}. Actually, $\rho_{12}^{out}$
in Eq.~(\ref{tn}) is the \textit{classical-quantum state}, as dubbed in
Ref.~\cite{disc-adPRA}---While for measurements on neuron-1 the discord is
zero, measurements on neuron-2 in general lead to non-zero discord.

Thus, we reveal the first important feature of our neural network (a soft
quantum computer). Namely, the neural network can develop quantum correlations
although only mixed initial states and very noisy operations are available.
Consequently, if soft quantum computing does mimic the working mechanism of
the human brain, the brain is certainly quantum as there are quantum
correlations among the neurons therein.

Now let us consider the three-neuron case. We are interested in two particular
examples, where two controlled Kraus operations $\mathcal{O}_{13}$ and
$\mathcal{O}_{23}$ are performed in different orders, namely, $\rho
_{123}^{out}=\mathcal{O}_{13}\mathcal{O}_{23}(\rho_{1}^{in}\otimes\rho
_{2}^{in}\otimes\rho_{3}^{in})$\ and\ $\rho_{123}^{\prime out}=\mathcal{O}%
_{23}\mathcal{O}_{13}(\rho_{1}^{in}\otimes\rho_{2}^{in}\otimes\rho_{3}^{in})$.
Assuming $\rho_{1}^{in}=p_{1}\left\vert 0\right\rangle _{1}\left\langle
0\right\vert +(1-p_{1})\left\vert 1\right\rangle _{1}\left\langle 1\right\vert
$\ and $\rho_{2}^{in}=p_{2}\left\vert 0\right\rangle _{2}\left\langle
0\right\vert +(1-p_{2})\left\vert 1\right\rangle _{2}\left\langle 1\right\vert
$\ yields\ %

\begin{align}
\rho_{123}^{out} &  =p_{1}(1-p_{2})\left\vert 0\right\rangle _{1}\left\langle
0\right\vert \otimes\left\vert 1\right\rangle _{2}\left\langle 1\right\vert
\otimes\mathcal{W}_{23}(\rho_{3}^{in})\nonumber\\
&  +p_{2}(1-p_{1})\left\vert 1\right\rangle _{1}\left\langle 1\right\vert
\otimes\left\vert 0\right\rangle _{2}\left\langle 0\right\vert \otimes
\mathcal{W}_{13}(\rho_{3}^{in})\nonumber\\
&  +p_{1}p_{2}\left\vert 0\right\rangle _{1}\left\langle 0\right\vert
\otimes\left\vert 0\right\rangle _{1}\left\langle 0\right\vert \otimes\rho
_{3}^{in}+(1-p_{1})\nonumber\\
&  \times(1-p_{2})\left\vert 1\right\rangle _{1}\left\langle 1\right\vert
\otimes\left\vert 1\right\rangle _{2}\left\langle 1\right\vert \otimes
\mathcal{W}_{13}\mathcal{W}_{23}(\rho_{3}^{in}).\label{threen}%
\end{align}
Meanwhile, it is easy to check that $\rho_{123}^{\prime out}$ can be obtained
from $\rho_{123}^{out}$ only by reversing the orders of $\mathcal{W}_{13}$ and
$\mathcal{W}_{23}$ in the last term of Eq.~(\ref{threen}). If $(1-p_{1}%
)(1-p_{2})\neq0$, then for noncommutative $\mathcal{W}_{13}$ and
$\mathcal{W}_{23}$\ the resulting states $\rho_{123}^{out}$ and $\rho
_{123}^{out}$ are different from each other. The importance of the quantum
operation orders is a quantum feature, which remains in our quantum neural
network model. Using the national representative surveys (Gallup polls) and
laboratory experiments, a recent result \cite{order-PNAS} reported an evidence
supporting a quantum model of question order effects for human judgments.
Whether or not our quantum neural network model is a quantum mechanical
foundation for such a result is certainly an interesting future problem.

Open quantum systems have been considered for various applications in quantum
information processing by engineering environment \cite{Zoller-NP,Zoller-N}
and in particular, for mixed-state quantum computing
\cite{onebit,ms-QC-1998,ms-QC-2002,ms-QC-2011} or dissipative quantum
computing \cite{VWCirac,dissiCT-theorem}. While soft quantum computing is no
more powerful than the unitary circuit model as implied by the
\textquotedblleft dissipative quantum Church-Turing theorem\textquotedblright%
\ \cite{dissiCT-theorem}, there does exist a mixed-state quantum computing
model (with a collection of qubits in the completely mixed state coupled to a
single control qubit with nonzero purity), proposed by Knill and Laflamme
\cite{onebit} and known as deterministic quantum computation with one qubit
(DQC1), which provides exponential speedup (for estimating the
normalized-trace of a unitary matrix) over the best known classical
algorithms. The possible role of quantum discord was considered, e.g., in
Refs.~\cite{Brukner,power1q,QC-disc}, as a figure of merit for characterizing
the nonclassical resources present in the DQC1, which itself is a special soft
quantum computer. Thus, it is reasonable to expect that soft quantum computing
has computational power between classical computation and usual quantum computation.

Finally, we give a brief remark on the physical implementation of soft quantum
computing. Above discussions focus mainly on a special form of the conditional
dynamical evolution as in Eq.~(\ref{evo}); more general forms can of course be
envisioned. A physical realization of $\mathcal{O}_{ij}$ is simple: The
superprojectors\ $\mathcal{P}_{\left\vert s\right\rangle _{i}}$ correspond to
a von Neumann measurement on neuron-$i$ along the computational basis;
conditionally on the measurement results, an identity operation $\hat{I}$ or
$\mathcal{W}_{ij}(t)$ is acted upon neuron-$j$. Such a conditional dynamical
evolution is experimentally friendly to implementation based on any quantum
computing systems under current investigation, such as linear optics,
superconducting qubits, atoms/ions, and quantum dots.

To summarize, we have proposed soft quantum computing as nonclassical
computation using real world quantum systems with naturally occurring
decoherence and dissipation induced by environment. A mathematical model of
quantum neural network is suggested to illustrate soft quantum computing. Even
using very simple and noisy operations [the controlled Kraus operations in
Eq.~(\ref{evo})], the quantum features, such as the presence of quantum
discord and non-commutability of quantum operations, remain in our model. As
an experimentally friendly model, the neural network as proposed mimics as
much as possible the realistic human brain and thus, paves the way to quantum
artificial intelligence\ and to better understanding the working mechanism of
the human brain. If the human brain does be a soft quantum computer, it
utilizes quantum laws in such a fundamental way that it forms certainly a
quantum correlated entity. This might be the reason why the human brain is
much more powerful for certain tasks than classical computers, although we
have to know more consequences and applications of the model to arrive at a
conclusive result. While the soft quantum computer is much easier to build
than the standard one, we need to do more works to understand what kinds of
computational tasks it can execute and to find the corresponding efficient algorithms.

The author gratefully acknowledges warm hospitality of the colleagues in
School of Physics, Nanjing University, where the present work was finished.

\end{document}